\documentclass[pre,showpacs,twocolumn,epsf]{revtex4}
\usepackage{graphicx} 
\usepackage{dcolumn}
\usepackage{pstricks}
\input epsf 

\def\openone{\leavevmode\hbox{\small1\kern-3.3pt\normalsize1}}
\newcommand{\la}{\langle}
\newcommand{\ra}{\rangle}
\newcommand{\be}{\begin{equation}}
\newcommand{\ee}{\end{equation}}
\newcommand{\bea}{\begin{eqnarray}}
\newcommand{\eea}{\end{eqnarray}}
\newcommand{\pd}{\partial}

\begin{document}

\title{Power--law tail distributions and nonergodicity}
\author{Eric Lutz}
\affiliation{Department of Quantum Physics, University of Ulm,
D-89069 Ulm, Germany\\
Sloane Physics Laboratory, Yale University, P.O. Box 208120, New Haven,
CT 06520-8120}
 
\date{\today}

\begin{abstract}
We  establish an explicit correspondence between ergodicity breaking in a system described by power--law tail distributions and the divergence of the  moments of these distributions.
\end{abstract}
  
\maketitle
Statistical mechanics is a  combination of the law of large numbers and the laws of mechanics. Since its foundation in the late 19th century, this theory has been extremely successful in describing equilibrium and  nonequilibrium properties of a very large number of  macroscopic systems \cite{dor99}.  In the last decade, however, a new class of systems that do not obey the law of  large numbers has emerged \cite{shl93,shl94,pek98}.  The behavior of these systems is  dominated by large and rare fluctuations that  are characterized by broad  distributions with power--law tails.
The hallmark  of these statistical  distributions, commonly referred to as L\'evy statistics \cite{lev37}, is the divergence of their first and/or second moment.

The question we  address in this paper is how  ergodicity is  affected in systems described by power--law tail distributions with diverging moments.  Ergodicity is a central concept in statistical physics and is usually stated  by saying that  ensemble average and time average of observables are equal in the infinite--time limit \cite{leb73,jan63}. The ergodic hypothesis has recently been investigated experimentally  in two different systems governed by L\'evy statistics in time: fluorescence intermittency of nanocrystal quantum dots \cite{bro03} and subrecoil laser cooling of atoms \cite{sau99}. Both experiments have  found that L\'evy statistics induces ergodicity breaking. A precise understanding of the connection between  divergent moments and nonergodicity is thus of high interest. A common feature of the  systems in the above experiments is their non--stationarity. This in in contrast to  the system we propose to study, namely atomic transport in an optical lattice, where a steady state  does exist. Nonetheless,  we will show that this  system can exhibit nonergodic behavior. 

An optical lattice is a standing wave light field formed at the intersection of two or more laser beams \cite{gry01}. Due to the spatial periodicity of the  potential, an optical lattice is  similar in many respects to a solid state crystal. The main advantage of an optical lattice, however,  is its high tunability: both the period and the amplitude of the optical potential can be modified in a controlled way. This has the  interesting consequence that  the exponents of the power--law distributions appearing in  this system are not fixed, as in most systems, but can be varied continuously, allowing the exploration of  different regimes.  

The motion of atoms  in a one-dimensional optical lattice formed by two counterpropagating laser beams with linear perpendicular polarization can be described, after spatial averaging,  by   a Fokker--Planck equation for the semiclassical Wigner function $W(p,t)$ 
\cite{cas91,hod95,mar96},
 \be
 \label{eq1}
 \frac{\pd W}{\pd t} =
 -\frac{\pd}{\pd p}\left[K(p)W\right] + \frac{\pd }{\pd
 p}\left[D(p)\frac{\pd W}{\pd p}\right] \ .
 \ee
The  momentum--dependent drift and diffusion
coefficients are respectively given by,
 \be \label{eq2} K(p) = -\frac{\alpha
 p}{1+(p/p_c)^2}, \hspace{0.3cm} D(p) = D_0 +
 \frac{D_1}{1+(p/p_c)^2} \ .
 \ee
 The drift $K(p)$ corresponds to a cooling
force  with friction coefficient
$\alpha$, while the diffusion coefficient $D(p)$ represents stochastic momentum fluctuations and
describes  heating processes. It is worth noticing that for large momentum, the range of the drift is limited by the capture momentum $p_c$, while the range of the fluctuations is not. The kinetic   equation (\ref{eq1}) is valid in a regime where (i) the atomic momentum is large,  $p\gg \hbar k$, where $k$
is the wave number of the laser field (this  defines the semiclassical limit), (ii) the
saturation parameter is small, $s\ll1 $ (this corresponds to low laser intensity) and (iii) the kinetic energy of the atoms is large, $p^2/2m \gg U_0$, where $U_0$ is the depth of the optical potential  (this last condition allows spatial averaging). The stationary solution of the atomic transport equation (\ref{eq1}) which satisfies natural boundary conditions, $W_s(p\rightarrow \pm \infty) \rightarrow 0$, is of the form,
\be
\label{eq3}
W_s(p) = \frac{1}{Z} \Big[ 1+ \frac{D_0}{D_0+D_1} \frac{p^2}{p_c^2}\Big]^{-(\alpha p_c^2)/(2D_0)} \ ,
\ee
where $Z$ is a normalization constant. The  momentum distribution  (\ref{eq3}) has an asymptotic power--law tail, $W_s(p) \sim |p|^{-(\alpha p_c^2)/D_0}$, with an exponent that can be  expressed in terms of the potential depth $U_0$ and the recoil energy $E_R$ as $(\alpha p_c^2)/D_0=U_0/(22 E_R)$ \cite{cas91}. The statistical properties of the distribution  (\ref{eq3}) can therefore be easily changed from normal statistics for $U_0\geq 66 E_R$, to L\'evy statistics for $U_0< 66 E_R$, by simply modifying the depth of the optical potential. In particular, one should note that the second moment, $\la p^2\ra =\int dp\,p^2 W_s(p)$, becomes infinite when $U_0\!<\!66 E_R$. In this regime the mean kinetic energy of the system, $E_K=\la p^2\ra/2m$, diverges, clearly signaling unusual thermodynamic behavior.

Transport in an optical lattice has a number of attractive features that make it an ideal case study of the thermodynamical properties of systems described by power--law distributions. On the one hand, the atomic transport equation (\ref{eq1}) has been derived from the  microscopic Hamiltonian that describes the interaction  with the laser fields and the quantities that appear in this equation can be expressed in terms of the microscopic parameters of the quantum--optical problem  \cite{cas91}. Moreover, the regime defined by conditions (i) to (iii) has been implemented experimentally  and the divergence of the kinetic energy below a given potential threshold has been observed \cite{kat97}. Finally,   Eq.~(\ref{eq1}) is an ordinary linear Fokker--Planck equation,  meaning that  standard methods of stochastic calculus can be used to analyse the problem. This is in contrast to most systems with power--law distributions that are often described by nonlinear or fractional kinetic  equations \cite{met00}. In particular, for the case of  vanishing $D_1$ (which we shall consider in the sequel \cite{rem1}),   Eq.~(\ref{eq1}) corresponds to a random process driven by {\it additive} Gaussian white noise. The fact that power--law fluctuations with infinite variance occur here in a system subjected solely to  additive noise is worth emphasizing. The physical mechanism which gives rise to divergent fluctuations in systems with multiplicative noise, where the noise strength is proportional to the stochastic variable, is well--known \cite{tak97}: it is based on a positive feedback loop that leads to  the  amplification of the noise as the value of the random variable increases. On the other hand, in the present situation, where the noise intensity is independent of the random variable, a different positive feedback mechanism is at work, based on the steady decrease to zero of the friction force as the value of the momentum increases.  The appearance of infinite momentum fluctuations  in this system is thus a striking illustration of the complex behavior  that  can result from the subtle interplay of the noise and the nonlinearity of the drift. In the following, we establish a correspondence between  the divergent moments of the power--law  distributions of  the system and nonergodic behavior.

We begin  by transforming the Fokker--Planck equation (\ref{eq1}) into a Schr\"odinger--like  equation by  writing $W(p,t) = W_s(p)^{1/2} \times\psi(p,t)$  \cite{ris89}. The function $\psi(p,t)$  satisfies the imaginary--time Schr\"odinger equation,
\be
\label{eq4}
-\frac{\pd \psi}{\pd t}= -D_0 \frac{\pd^2 \psi}{\pd p^2} + V(p) \psi = H \psi \ ,
\ee
with the potential  $V(p) = K'(p)/2 + K(p)^2/(4D_0)$. For the drift coefficient (\ref{eq2}),  this  potential reads
\be
\label{eq5}
V(p) = \frac{\alpha p_c^2}{4D_0} \,\,\frac{ p^2 \,(\alpha p_c^2+ 2D_0)- 2D_0\, p_c^2}{(p^2+p_c^2)^2} \ .
\ee
This transformation reveals the fundamental difference between  the case of finite $p_c$ and the case of infinite $p_c$, where   Eq.~(\ref{eq1})  reduces to the familiar Ornstein-Uhlenbeck process with linear drift. 
For infinite $p_c$, the Schr\"odinger potential $V(p)$ asymptotically increases as $p^2$, whereas for finite $p_c$, it asymptotically {\it decreases} as $1/p^{2}$. By contrast, the Fokker--Planck potential, $\Phi(p) =-\int_0^p dp'\, K(p') $, is  confining for any value of the capture momentum. As a consequence, the spectrum of the Hamiltonian, $H \psi_k(p) = E_k \psi_k(p)$, is discrete in the former case, while it is continuous, except for the discrete ground state $\psi_0(p)$, in the latter. In both cases, the stationary momentum distribution is given by the square of the ground state eigenfunction, $W_s(p) = \psi_0(p)^2$. Interestingly, we note that a real--time Schr\"odinger equation with a potential of the form  (\ref{eq5}) has recently been considered in Ref.~\cite{lil00}.
To simplify the discussion, we now adopt rescaled variables for which $\alpha=p_c=1$ and $D_0=D$; the noise intensity $D$ being then the only remaining parameter  in the problem. 
It will also be convenient to divide momentum space into  a low--momentum region $|p|<1$, where the drift is approximately linear, $K_1(p) \simeq -p$, and a high--momentum region $|p|>1$, with a drift $K_2(p)  \simeq -1/p$. The anomalous dynamics of the system is completely determined by the high--momentum region. The eigenvalues and eigenfunctions of the Hamiltonian  $H$  in this region  are given by $E_k= D k^2$ and $\psi_k(p) = \sqrt{p}\, [ c_1 J_\nu(k p)+ c_2 Y_\nu(k p)]$, where $J_\nu(p)$ and $Y_\nu(p)$ are  the Bessel functions of the first and second kind of order $\nu = (D+1)/2D$. The constants $c_{1,2}$ are fixed by  the boundary   conditions. 

When discussing the ergodicity of a system, one is typically  not interested in the trajectory in the full space--space, but often in the projection of the trajectory onto  some subspace of relevant variables \cite{dor99}, in the present case momentum. A  criterion for the equality of ensemble average and time average  of a dynamical quantity $A$ is then provided by the condition \cite{pap91}
\be
\label{eq6}
\sigma_A^2(t) = \la \Big(\overline{A\,}- \la \overline{A\,}\ra   \Big)^2\ra \longrightarrow 0\mbox{ when } t\longrightarrow \infty\ .
\ee
Here $\overline{A\,} = t^{-1} \int_0^t d\tau \,A(p(\tau))$ is the time average of the observable $A$  and $\la A\ra = \int dp\,  A(p) W(p,t)$ denotes its ensemble average. In the infinite--time limit, the latter  tends to the stationary ensemble average $\la A\ra_s = \int dp\,  A(p) W_s(p)$. A system that obeys (\ref{eq6}) is said to be ergodic in the mean square sense.  In order to determine  the ergodicity of our system, we thus need to compute the long--time behavior of the  covariance,
\bea
\label{eq7}
\sigma_A^2(t)= \frac{1}{t^2}\int_0^t dt_1 \!\int_0^t dt_2 \,\Big [&&\!\!\!\!\!\la A(p(t_1))A(p(t_2))\ra  -\nonumber \\ &&\!\!\!\!\!\la  A(p(t_1))\ra \,\la A(p(t_2))\ra \Big] .
\eea
This can be done by applying the usual theory of stochastic processes \cite{ris89}.
The two--time correlation function $ \la A(p(t_1))A(p(t_2))\ra$  is defined by the integral
\bea
\label{eq8}
\la A(p(t_1))\!\!&&\!\!\!\!\!A(p(t_2))\ra =\nonumber \\
&&\!\!\!\!\!\int\! \!\int\! dp_1 dp_2 \, A(p1)A(p_2) \,W_2(p_1,t_1; p_2,t_2)\ ,
\eea
where $W_2(p_1,t_1; p_2,t_2)$ is the two--point joint probability density function. Since the Fokker--Planck equation (\ref{eq1}) describes a stationary Markov process (for any value of $p_c$), the probability distribution $W_2(p_1,t_1; p_2,t_2)$ depends only on the time difference $|t_1-t_2|$ and  can be expressed in terms of  the eigenvalues and eigenfunctions of the Schr\"odinger equation (\ref{eq4}) in the form,
\bea
\label{eq9}
&&\!\!\!\!\!W_2(p_1,t_1;p_2,t_2)= \psi_0(p_1) \psi_0(p_2) \times \nonumber \\
&&\hspace{-.7cm}\Big ( \psi_0(p_1) \psi_0(p_2) + \int_0^\infty dk \,\, \psi_k(p_1) \psi_k(p_2) \,e^{-E_k |t_1-t_2|}\Big)\ .
\eea
Combining Eqs.~(\ref{eq7}),  (\ref{eq8}) and (\ref{eq9}) and introducing the variable $\tau= t_2-t_1$, we arrive at
\be
\label{eq10}
\sigma_A^2(t) =  \frac{2}{t^2}\int_0^t d\tau \, (t-\tau) \,C_A(\tau)\ , 
\ee
where the function $C_A(\tau) $ is given by
\be
C_A(\tau) = \!\int_0^\infty dk\, \Big [\int \!dp\, A(p) \, \psi_0(p) \psi_k(p)\Big]^2 \,e^{-E_k \tau}  \ .
\ee
The infinite--time limit of the covariance $\sigma_A^2(t) $ is entirely determined by the asymptotic behavior of  $C_A(\tau)$. This function depends explicitely on the observable $A(p)$. It is  worthwhile  to notice  that ergodicity  will therefore in general depend on the dynamical variable $A(p)$ under consideration. In the following, we take $A(p) = p^n$ and evaluate the long-time behavior of  $C_A(\tau)$ following Ref.~\cite{mar96}. We find $C_A(\tau)\!\sim\! \tau^{-\mu}$ with an exponent $\mu = (1-(2n+1) D)/2D$. As a result, the covariance (\ref{eq10}) will converge to zero as $t\rightarrow \infty$, only if $D<D_n = 1/(2n+1)$. We thus obtain the important result that there is  a noise threshold $D_n$ above which ergodicity is broken. As already mentioned, this threshold depends explicitely on the parameter $n$, that is, on the quantity $A(p)$: the smaller the power $n$, the larger the value of $D_n$. On the other hand, the moments of the stationary momentum distribution, $\la p^m\ra = \int dp\, p^m W_s(p)$, are finite for $D<D'_m=1/(m+1)$. We can therefore  conclude that there exists  a direct relationship between the loss of ergodicity in the system for  the variable $A(p) = p^n$ and the divergence of the $2n$th moment of the stationary momentum distribution.

Let us  look in more detail at the value $n=0$, which corresponds to the largest noise threshold $D_{n=0} = 1$. We first mention that for $D\!>\!D_{n=0}$, the Fokker--Planck equation doesn't have  a normalizable stationary solution anymore and the system is obviously nonergodic. Further, for  $n=0$, the function $C_A(\tau)$ can be rewritten in terms of the conditional probability density  $P_2(p_2,\tau|p_1,0)$ as 
\bea
\label{eq11}
C_A(\tau) = &&\int dp_1\, W_s(p_1) \times\nonumber \\
&&\int dp_2\,  \Big( P_2(p_2,\tau|p_1,0) -W_s(p_2)\Big) \ .
\eea
This result is of special interest. Equation (\ref{eq11}) shows that, unlike for the Ornstein--Uhlenbeck process, the steady  state is here  reached in a nonexponential way.  Such an asymptotic power--law decay is usually associated  with  non--Markovian processes and  is rather surprising for  a stationary Markov process as described by  Eq.~(\ref{eq1}). The origin of  this algebraic behavior is of course rooted in the nonlinear drift coefficient  (\ref{eq2}). 
\begin{center}
\begin{figure}[t]
\epsfxsize=0.3\textwidth
\epsffile{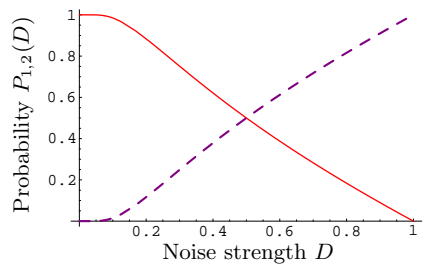}
\caption{Probability to be in the low--momentum region (continuous line) and in the high--momentum region (dashed line), in the long--time limit, as a function of the noise strength $D$.}
\label{fig}
\end{figure}
\end{center}
\vskip -1cm

A closely related quantity is the first--passage time distribution. The first--passage time is defined as the time at which the momentum of the system first exits a certain momentum interval, given that it was originally in that interval. The first--passage time problem for the Fokker--Planck equation (\ref{eq1}) can  again be treated using standard techniques of stochastic calculus (see for example Ref.~\cite{pra65}). In the high--momentum region, the Laplace transform $g_2(s)$ of the first--passage time distribution  obeys the following backward equation,
\be
\label{eq13}
D \, \frac{\pd^2 g_2}{\pd p^2} - \frac{1}{p} \,\frac{\pd g_2}{\pd p}- s  g_2=0\ .
\ee 
Solving Eq.~(\ref{eq13}) with the boundary conditions $g(1)=1$ and $g( \infty)=0$, we obtain
\be
\label{eq14}
g_2(s) = \frac{K_\nu(p\, \sqrt{s/D} )}{K_\nu(\sqrt{s/D})} \, p^\nu \ , 
\ee
where $K_\nu(p) $ is the modified Bessel function of the second kind of order $\nu$. It follows from  Eq.~(\ref{eq14}), that to leading order, $g_2(s) \sim s^\nu$ as $s\rightarrow 0$. The first--passage time distribution in region 2 is thus also a power--law tail distribution and it  asymptotically behaves as $g_2(t) \sim t^{-\gamma}$ with an exponent $\gamma={(3D+1)/(2D)}$. The  corresponding  moments $\la t^n\ra = \int dt\, t^n g_2(t) $  converge if  $D<D''_n=1/(2n-1)$.  The first--passage time distribution $g_1(t)$ in  region 1 can be computed along similar lines and the associated moments can be shown to be all  finite.  Figure (\ref{fig}) shows the probability to be in region 1 and 2, in the limit of long times, as a function of the noise intensity $D$.  We  observe that for small $D$, atoms are mostly located in the low--momentum region, where they experience the linear part of the drift. On the other hand, for $D$ close to $D_{n=0}$, atoms get localized in the high--momentum region, where the drift asymptotically decays as $-1/p$. Remarquably, atoms in this system can  thus be  brought in a high--energy state through the sole action of the noise.  
We note that this problem exhibits an interesting analogy with subrecoil laser cooling, where atoms accumulate in a low--energy state (in a so--called dark state), with infinite mean trapping time \cite{bar02}. To our knowledge, loss of ergodicity in systems with divergent trapping times has been first discussed in the context of spin--glasses \cite{bou92}.

We can now formulate the main result of the paper: Ensemble  average and time average of  the dynamical variable $A(p) = p^n$ stop being  equal in the infinite--time limit --- ergodicity of the system is accordingly broken---  when the $2n$th moment of the stationary momentum distribution and the $(n+1)$th moment of the first passage time distribution in the high--momentum region become infinite. An unambiguous  correspondence between the nonergodic properties  of a system described power--law  distributions and the divergence of their respective moments is therefore demonstrated. This confirms and  extends the findings reported in Refs.~\cite{sau99,bro03,bar02,bou92}. We stress that the above result is not restricted to transport in an optical lattice and that our  analysis applies to a whole class of systems described by an equation of the form (\ref{eq1}) with a drift coefficient decaying asymptotically as $-1/p$. More  generally, we conjecture  that the result also holds true for other power--law systems, even if they  are not described by a simple kinetic equation like the ordinary Fokker--Planck equation (\ref{eq1}) \cite{sau99,bro03,bar02,bou92}. 
   
Further insight can be  gained by  considering  a discretized form of the Fokker--Planck equation (\ref{eq1}). In doing so,  we shall obtain  a generalization of the Ehrenfest urn model, which  has played an important role in clarifying the foundations of statistical mechanics  \cite{kac54}. We write  $p\!=\!j \Delta p$ and $t\!=\!l \Delta t$ and find that the probability $\omega(j,l)=W(j \Delta p, l \Delta t) $ satisfies the  difference equation, 
\be
\label{eq15}
\omega(j,l+1) = a(j-1)\, \omega(j-1,l)+b(j+1) \, \omega(j+1,l) \ .
\ee  
In the  limit $(\Delta p, \Delta t)\! \rightarrow\! 0$,  Eq.~(\ref{eq15}) reduces   to the continuous equation (\ref{eq1}) with $K(p) \!=\! \Delta p/\Delta t\,\, [a(p)-b(p)]$ and $D\! =\! \ (\Delta p)^2/(2\Delta t)\, [a(p)+b(p)]$. The transition probabilities $a(j)$ and $b(j)$ in (\ref{eq15}) are explicitely given by
\vskip -.5cm
\be
\label{eq16}
a(j) = \frac{R(1+j^2) -j}{2R (1+j^2)}, \hspace{0.3cm}b(j) =\frac{R(1+j^2) +j}{2R (1+j^2)} \ .
\ee
\vskip -.1cm
\noindent
We recall that the Ehrenfest model consists of two urns containing a total of $2R$ balls. At regular time intervals, $\Delta t$, a ball is randomly chosen and moved to the other urn;  $w(j,l) $ is  then the probability of having $R+j$ balls in the first urn at time $l$.  In the standard version of the model, the $j^2$ terms in Eq.~(\ref{eq16}) are absent and Eq.~(\ref{eq1}) reduces to the Ornstein--Uhlenbeck process with linear drift. We note that in the continuous description, the momentum $p$ corresponds to the number of excess balls $j$ in the first urn. So, for example, ergodicity breaking for $n\!=\!1$ ($D\!>\!1/3$) occurs when the fluctuations of the number of balls in each urn diverge, the average number of balls being still finite and equal. Moreover, when $D$ approaches one, all the balls preferentially occupy the same urn, therefore  acting  as Maxwell's demon.  

In conclusion, we have investigated anomalous transport in an optical lattice from the point of view of  statistical mechanics  and established an explicit correspondence between ergodicity breaking and the divergence of the moments of the power--law tail distributions describing the behavior of the system, both in momentum space and in time.

We would like to thank F. Bardou for stimulating discussions. This work was funded in part  by the ONR under contract N00014-01-1-0594.


\vskip -0.55cm
\begin{thebibliography}{99}
\bibitem{dor99} J.R.  Dorfman, {\it An Introduction to Chaos in Nonequilibrium Statistical Mechanics} (Cambridge University Press, Cambridge, 1999).
\bibitem{shl93} M.F.  Shlesinger, G.M.  Zaslavsky, and J.  Klafter,
Nature {\bf 363}, 31 (1993).
\bibitem{shl94}
{\it L\'evy Flights and Related Topics in Physics},
  edited by M.F. Shlesinger, G.M.  Zaslavsky, and U.  Frisch, (Springer, Berlin, 1994).
\bibitem{pek98} {\it Anomalous Diffusion. From Basics to
    Applications}, edited by A.  P\c{e}kalski and K.  Sznajd--Weron,  
   (Springer, Berlin, 1998).
\bibitem{lev37}
 P. \L\'evy, {\it Th\'eorie de l'Addition des Variables Al\'eatoires}, (Gauthiers--Villars, Paris, 1937).
\bibitem{leb73} J.L. Lebowitz and O.  Penrose,  Physics Today  {\bf 26} (no. 2) 23 (1973).
\bibitem{jan63} R. Jancel, {\it Foundations of Classical and Quantum Statistical Mechanics}, (Pergamon Press, Oxford, 1963).
\bibitem{bro03} X. Brokmann {\it et al.}, Phys. Rev. Lett. {\bf 90}, 120601 (2003).
\bibitem{sau99} B. Saubamea, M. Leduc, and C.  Cohen-Tannoudji,
Phys. Rev. Lett. {\bf 83}, 3796 (1999).
\bibitem{gry01} G. Grynberg and C. Robilliard, Phys. Rep. {\bf 355}, 335 (2001).
\bibitem{cas91} Y. Castin, J. Dalibard, and C. Cohen--Tannoudji, in {\it Light Induced Kinetic Effects on Atoms, Ions and Molecules}, edited by L. Moi {\it et al.}, (ETS Editrice, Pisa, 1991).
\bibitem{hod95} T.W. Hodapp {\it et al.}, Appl. Phys. B {\bf 60}, 135 (1995).
\bibitem{mar96} S. Marksteiner, K. Ellinger, and P. Zoller, Phys. Rev. A {\bf 53}, 3409 (1996).
\bibitem{kat97} H. Katori, S. Schlipf, and H. Walther, Phys. Rev. Lett. {\bf 79}, 2221 (1997).
\bibitem{met00} R. Metzler and J. Klafter, Phys. Rep. {\bf 339}, 1 (2000). 
\bibitem{rem1} This is  justified since a) the exponent in (\ref{eq3}) in independent of $D_1$ and b)  the term proportional to  $D_1$ gives a negligible contribution to the noise in the limit of  large \nolinebreak $p$.
\bibitem{tak97} H. Takayasu, A.-H. Sato, and M. Takayasu, Phys. Rev. Lett. {\bf 79}, 966 (1997). 
\bibitem{ris89} H. Risken, {\it The Fokker-Planck Equation}, (Springer, Berlin, 1989).
\bibitem{lil00} F. Lillo and R.N. Mantegna, Phys. Rev. Lett. {\bf
84}, 1061 (2000); {\bf 84}, 4516 (2000).
\bibitem{pap91} A. Papoulis, {\it Probability, Random Variable, and Stochastic Process}, (McGraw Hills, New York, 1991).
\bibitem{pra65} N. U. Prabhu, {\it Stochastic Processes}, (Macmillan, New York, 1965).
\bibitem{bar02} F. Bardou {\it et al.}, {\it L\'evy Statistics and Laser
Cooling} (Cambridge University Press, Cambridge, 2002).
\bibitem{bou92} J.-P. Bouchaud, J. Phys. I (France) {\bf 2}, 1705 (1992).
\bibitem{kac54} M. Kac,  in N. Wax (Ed.), {\it Selected Papers on Noise and  Stochastic Processes} (Dover, New York, 1954).
\end{thebibliography}
\end{document}